\documentclass[aps,prx,twocolumn,longbibliography,floatfix,superscriptaddress]{revtex4-1}

\usepackage{graphicx}
\usepackage{epstopdf}
\usepackage{epsfig}
\usepackage{amssymb,amsmath,stmaryrd,tabularx}
\usepackage{wrapfig}
\usepackage{bm}
\usepackage[T1]{fontenc}
\usepackage[center]{subfigure}
\usepackage[toc,page]{appendix}
\usepackage{float}
\usepackage[percent]{overpic}
\usepackage[dvipsnames]{xcolor}

\renewcommand{\d}{\partial}
\renewcommand{\L}{\mathcal L}
\newcommand{\e}{\epsilon}

\renewcommand{\a}{\alpha}

\newcommand{\ext}{\textrm{ext}}
\let\vec\mathbf

\begin{document}
\title{Dislocation nucleation in the phase field crystal model} 

\author{Vidar Skogvoll}
\affiliation{PoreLab, The Njord Centre, Department of Physics, University of Oslo, P. O. Box 1048, 0316 Oslo, Norway}
\author{Audun Skaugen}
\affiliation{Computational Physics Laboratory, Tampere University, P.O. Box 692, FI-33014 Tampere, Finland}
\author{Luiza Angheluta}
\affiliation{PoreLab, The Njord Centre, Department of Physics, University of Oslo, P. O. Box 1048, 0316 Oslo, Norway}
\author{Jorge Vi\~nals}
\affiliation{School of Physics and Astronomy, University of Minnesota, Minneapolis, MN 55455
}

\begin{abstract} 
We use the phase field crystal model to study nucleation of edge dislocations in two dimensions under an applied stress field. A dislocation dipole nucleates under the applied stress, consistent with Burgers vector conservation. The phase field correctly accounts for elastic energy storage prior to nucleation, and for dissipative relaxation during the nucleation event. We show that a lattice incompatibility field is a sensitive diagnostic of the location of the nucleation event, and of the Burgers vector and slip direction of the dislocations that will be nucleated above threshold. A direct calculation of the phase field energy accurately correlates with the nucleation event, as signaled by the lattice incompatibility field. We show that a Schmid-like criterion concerning the resolved stress at the nucleation site correctly predicts the critical nucleation stress. Finally, we present preliminary results for a three-dimensional, body-centered cubic lattice. The phase field allows a direct computation of the lattice incompatibility tensor for both dislocation lines and loops.
\end{abstract}

\maketitle


\section{Introduction}
Unlike the spontaneous homogeneous nucleation of topological defects in a symmetry-breaking phase transition \cite{re:langer69,re:gunton83,re:davis95}, the formation of dislocation lines in a material is typically studied as an athermal process largely driven by local stresses \cite{FriedelDislocationIntroduction1979}. Since the existence and mobility of such defects are essential contributors to the strength and ductility of crystalline materials, understanding the mechanisms behind their creation and motion is a fundamental goal of material science in general, and of plasticity theory in particular. Along parallel developments in the continuum theory of crystal plasticity, a number of empirical criteria have been introduced to predict dislocation nucleation thresholds, the resulting Burgers vector distribution, and line direction \cite{re:phillips01,re:li02,miller2004stress,re:garg15}. These macroscopic criteria have been extensively compared with microscopic results from Molecular Dynamics (MD) simulations of model crystalline solids in a variety of configurations and imposed stresses \cite{re:kelchner98,re:zhu08,re:li10,re:garg15}. However, the details of the mechanical conditions that lead to dislocation nucleation still remain poorly understood, with criteria from continuum mechanics approaches and numerical simulations often yielding conflicting phenomenology. The two main reasons why a precise comparison between the two is difficult include the disparity in length scales between crystal plasticity theory and molecular simulation, and the necessity in the latter to thermally average phase space trajectories that take place over characteristic energy scales which are much higher than thermal scales. Fundamental questions such as whether the nucleation event is local or nonlocal, remain unresolved \cite{re:garg15}. We bridge here microscopic and continuum scales by introducing a phase field crystal model \cite{re:elder02,re:emmerich12} of dislocation nucleation, and show that the nucleation event is well captured at the mesoscale by a continuum lattice incompatibility field. Our numerical results for the nucleation of edge dislocations in a two dimensional (2D), hexagonal lattice indicate that the nucleation event is governed by a local balance between the resolved stresses along lattice slip planes and the force acting between the nucleating dislocation pair, and that a lattice incompatibility field derived from the phase field predicts the Burgers vector of the nucleating defect pair.
The simplest dislocation nucleation criterion is based on the Schmid stress decomposition \cite{rice1992dislocation,re:phillips01,li2004elastic,miller2004stress,miller2008nonlocal,garg2016mechanical}. When an appropriate projection of an atomic level shear stress exceeds a material dependent threshold, a dislocation loop is predicted to be nucleated. On the one hand, while fcc lattices generally obey the Schmid criterion, there exist entire classes on \lq\lq non-Schmid" lattices, including bcc crystals \cite{re:duesbery98}. Furthermore, a recent, careful MD study of nucleation in a nanoindentation configuration for a model Lennard-Jones solid shows that the Schmid criterion not only fails to account for the site of the nucleation event, but that nucleation, in fact, occurs in regions in which the resolved shear stress is relatively small \cite{re:garg15}. A second class of criteria associate the nucleation event to a buckling or phonon instability of the lattice (the Hill or $\Lambda$ criteria based on mechanical stability arguments \cite{miller2004stress,miller2008nonlocal}). Molecular Dynamics simulations and experiments in different crystal indentation configurations, however, have revealed very complex nucleation processes in which the lattice is locally quite distorted, and therefore far from the conditions of applicability of such a phonon stability analysis. Large regions of partial dislocations and extended stacking faults have been argued to be present at nucleation \cite{re:kelchner98}, as well as extended and complex networks involving surfaces and grain boundaries \cite{re:li10}. More recently, the stability of the perfect lattice against homogeneous nucleation has been formulated in terms of the kinematic equation that governs the temporal evolution of the dislocation density tensor. This approach is sensitive to the creation of non-trivial local topology \cite{re:garg15}, and yields predictions that are qualitatively different than the Schmid criterion. To contribute to the elucidation of the criteria for nucleation, we examine here a simple, prototypical configuration: A 2D, hexagonal lattice, in which nucleation occurs through the formation of a dislocation dipole of zero net Burgers vector. In this idealized configuration, we show that the incompatibility field directly computed from the phase field identifies the nucleation event, and that it can be used to predict the Burgers vector at nucleation. The critical stress for nucleation is seen to be in quantitative agreement with that of the Schmid criterion in this 2D lattice.

\section{The phase-field crystal}
The phase field crystal (PFC) model is a mesoscale description of a crystalline solid in which vibrational degrees of freedom have been averaged out, in the same spirit as density functional theory \cite{elder2007phase,re:emmerich12}. The crystalline phase is described by a scalar order parameter field $\psi(\mathbf{r})$, which obeys a phenomenological free energy given, in dimensionless form, by 
\begin{equation}
\mathcal F[\psi] = \int d \vec r \left [ \frac{1}{2} [\L \psi]^2 + \frac{r}{2} \psi^2 + \frac{1}{4} \psi^4 \right ],
\label{eq:free_energy_of_PFC_model}
\end{equation}
where $\L = 1+\nabla^2$, and $r$ is a dimensionless parameter representing the deviation from the liquid-solid phase boundary. 
%
\begin{figure}[t]
\centering
\includegraphics[width=0.45\textwidth]{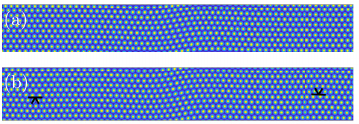} 
 \caption{Central region of the computational domain with (a), PFC configuration in equilibrium at $\sigma_0=0.080$ prior to nucleation, and (b), the equilibrium configuration at $\sigma_0=0.081$ after nucleation.}
     \label{fig:PFC_during_nucleation}
\end{figure}
%
We further assume that $\psi$ is a conserved variable with its spatial average  $\bar \psi$ being constant \cite{elder2007phase}. The two constants $r$ and $\bar \psi$ completely define the model. In 2D, for a range of values of $\bar \psi$ and $r<0$, a triangular Bravais lattice $\psi^{eq}(\vec r)$ is the equilibrium phase, which we consider here.

Upon deformation of the equilibrium state $\psi^{eq}$ by a displacement field $\vec u( \vec x )$, it is possible to define the equilibrium stress tensor as the variation of the free energy with respect to the displacement gradient \cite{skaugen2018dislocation},
 \begin{equation}
 \tilde\sigma_{ij}^\psi =
-  \d_m [\psi \d_m \L \psi ] \delta_{ij}
+ 2 [\d_{(i} \L \psi] [\d_{j)} \psi],
\label{eq:stress_tensor}
\end{equation}
 where $X_{(i} Y_{j)}$ refers to symmetrization of indices $i,j$ \cite{misc1}. This quantity still shows spatial variations within a unit cell due to the variation of $\psi$. We therefore further define an averaged stress field as $ \sigma_{ij}^\psi = \langle \tilde \sigma_{ij}^\psi \rangle $, where $\langle \cdot \rangle$ refers to averaging over an area approximately equal to a unit lattice cell.

For small distortions, the hexagonal lattice is elastically isotropic. We define a symmetric strain as
\begin{equation}
e_{ij}^\psi = \frac{1}{2\mu}(\sigma^\psi_{ij} - \kappa \delta_{ij} \sigma^\psi_{kk}).
\label{eq:strain_from_stress}
\end{equation}
where $\mu$ is the shear modulus, and $\kappa = \lambda /(2(\lambda+\mu))$, where $\lambda$ is the standard Lam\'e coefficient. In our dimensionless variables, we have $\lambda = \mu = 3 A_{0}^{2}$, where $A_{0}$ is the amplitude of $\psi^{eq}$ \cite{skaugen2018dislocation}. 

Dislocations lead to lattice incompatibility \cite{FriedelDislocationIntroduction1979,re:kronerContinuumTheoryDefects1981}. In 2D, and given a Burgers vector density $\vec B(\vec r)$, the incompatibility field is $ \eta = \e_{ik} \e_{jl} \d_{ij}  e_{kl}  = \epsilon_{ij}\partial_i B_j $. A key assumption of our analysis is that the configuration of $\psi$ contains the complete strain incompatibility \cite{skaugenSeparationElasticPlastic2018,re:acharya20}. Thus, from Eq.~(\ref{eq:strain_from_stress}) we find,
\begin{equation}
\eta^\psi = \frac{1}{2\mu} (\e_{ik} \e_{jl} \d_{ij} \sigma_{kl}^\psi-\kappa \nabla^2 \sigma_{kk}^\psi).
\label{eq:eta_sigma_definition}
\end{equation}

The dissipative evolution of $\psi$ is diffusive 
\begin{equation}
 \d_t \psi = \nabla^2 \frac{\delta \mathcal F}{\delta \psi},
\end{equation}
with a constant kinetic mobility coefficient, which we set to one in our study and which sets the unit of time. 
As discussed in Ref.~\cite{skaugenSeparationElasticPlastic2018,Heinonen_2016}, lattice distortion needs to be treated separately from diffusive relaxation of $\psi$ in order to incorporate elastic response into the phase field, and to maintain elastic equilibrium at all times. In addition, and in order to induce nucleation, we consider an externally imposed bulk stress $\sigma_{ij}^{ext}(\mathbf r)$. In elastic equilibrium $\d_i \sigma^\psi_{ij} = \d_i \sigma_{ij}^{ext} $. Following Ref.~\cite{skaugenSeparationElasticPlastic2018}, for a non-equilibrium configuration of $\psi$, we solve $ \d_i( \sigma_{ij}^{\psi} -\sigma_{ij}^{ext} + \sigma_{ij}^\delta) = 0$,  where $\sigma_{ij}^\delta =\lambda e_{kk}^\delta + 2\mu e_{ij}^\delta$, and $e_{ij}^{\delta}$ is a compatible strain $e_{ij}^\delta = (\partial_i u^\delta_j+\partial_j u^\delta_i)/2$. Diffusion of $\psi$ is supplemented at each time by distortion $ \psi(\vec r) \rightarrow \psi(\vec r - \vec u^\delta)$.
%
\begin{figure}
    \centering
    \includegraphics[width=0.45\textwidth]{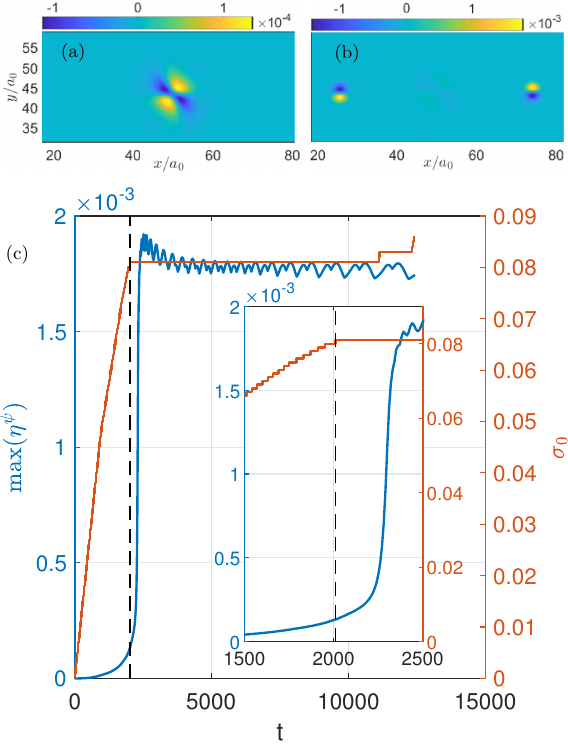} 
   \caption{a) Incompatibility field $\eta^\psi$ (a) at $t = 2050$ ($\sigma_0=0.080$), before the nucleation event, and (b) at $t = 12170$ ($\sigma_0=0.081$) after nucleation.  c) Maximum of $\eta^\psi$ as a function of time $t$. The dashed line at $t=2050$ marks where $\sigma_0$ has attained the critical value for nucleation. The dislocations become distinct at  $t \approx 2300$.
   The right axis shows the value of $\sigma_0$ for the corresponding times. The plateaus in time indicate non-equilibrium relaxation at constant external stress. }
    \label{fig:eta_psi_field_during_nucleation}
\end{figure}
%
In 2D, the condition for elastic equilibrium means that the  stress tensor difference can be written in terms of an Airy potential $\chi$, $\sigma_{ij}^\psi - \sigma_{ij}^{ext} + \sigma_{ij}^\delta = \e_{ik}\e_{jl} \d_{kl} \chi$. For each instantaneous configuration of $\psi$ we solve \cite{skaugenSeparationElasticPlastic2018},
\begin{equation}
\frac{1-\kappa}{2\mu} \nabla^4 \chi = \eta^\psi - \eta^{ext},
\end{equation}
where $\eta^{ext}$ accounts for the fact that the imposed stress does not necessarily derive from a compatible displacement. The solution allows the computation of $e_{ij}^{\delta}$ and, from it, of the displacement $u_{i}^{\delta}$.

\section{Numerical Method}

A square computational domain is considered with periodic boundary conditions, with $100 \times 100$ hexagonal unit cells of length $a_0=\frac{4\pi}{\sqrt 3}$, and grid spacings $\Delta x = a_0/7$ and $\Delta y = a_0 \sqrt 3/12$. The model parameters used are $r=-1$ and $\bar \psi=-0.45$. The initial condition of $\psi$ is a periodic, undistorted hexagonal lattice. For our choice of model parameters, the corresponding Lam\'e coefficients are $\mu = \lambda = 0.227$. 

Calibrating the parameter values to experiments is a difficult task due to a lack of high-resolution data and corresponding measurable quantities. Since the PFC free energy is an effective coarse-graining of the intermolecular potential as related to high-order density gradients, a substantial amount of fitting is required, beginning with energy scales, but also relaxation time scales \cite{emmerich2012phase}). Hence, the strength of the PFC model is not in modelling with specific dimensional units, but rather in modelling generic behavior described by rescaled units. To this end, the shear modulus $\mu$ sets the unit for measuring stress, while strain and incompatibility fields are dimensionless. For instance, the critical stress of $0.081$ in Fig. \ref{fig:eta_psi_field_during_nucleation} would correspond in physical units to to $\sigma^{*}_{c} = 0.081/0.227 \mu \approx 0.36 \mu$.

We impose a shear stress $\sigma_{xx}^{ext} = \sigma_{yy}^{ext}=0$ and $ \sigma_{xy}^{ext} = \sigma_0 e^{-\frac{|\vec r-\vec r_0|^2}{2w^2}},$ with $ \vec r_{0}$ an arbitrary center. Nucleation is induced by a sequence of steps of increasing value of $\sigma_{0}$. A given  configuration is allowed to relax to equilibrium for constant $\sigma_{0}$. After equilibration has been achieved, the configuration is used as the initial condition for another relaxation step in which the value of $\sigma_{0}$ is increased. The details are as follows: Diffusive relaxation of $\psi$ is allowed for $100$ steps by using an exponential time differencing method, with a time step of $\Delta t=0.1$ \cite{coxExponentialTimeDifferencing2002}. After these one hundred steps, $\psi$ is brought to mechanical equilibrium by a compatible displacement as described above, and in Refs. \cite{skaugenSeparationElasticPlastic2018,PFCamplitudesMarco20}. Diffusive relaxation and elastic distortion cycles are continued until the largest change in $\psi$ between two such cycles is less than $0.01$. We then increase $\sigma_0$ by an increment $\Delta \sigma_0=0.001$ and repeat the relaxation procedure. The external stress amplitude considered ranges from zero to $\sigma_0 = 0.086$.
Fig. \ref{fig:PFC_during_nucleation} shows the equilibrated field $\psi$ for some amplitude of $\sigma_{0}$ prior to ($\sigma_0 = 0.080$) and after a nucleation event ($\sigma_0 = 0.081$) for $w=4a_0$. The nucleation event gives rise to two edge dislocations with opposite Burgers vectors $a_0 \vec e_x$ and $-a_0 \vec e_x$. When the configuration comprising two defects is allowed to evolve, the defects move away from each other along $x$-direction in the figure. Note that, since a dislocation in a hexagonal lattice has two extra half-planes, we represent the location of the dislocation by the symbol $\veebar$. This is in contrast with the conventional symbol representing an edge dislocation ($\perp$) which indicates the directions of the slip and extra inserted half plane. A video animation of the nucleation event sequence can be seen in the Supplementary Material.

\section{2D Dislocation nucleation}

The incompatibility field $\eta^\psi$ from Eq. (\ref{eq:eta_sigma_definition}) accurately indicates where dislocations form. Fig. \ref{fig:eta_psi_field_during_nucleation}(a-b) shows the $\eta^\psi$ field corresponding to the $\psi$ density field in Fig.~\ref{fig:PFC_during_nucleation}, before and after the nucleation event. The extremes in the value of $\eta^\psi$ identify the location of the defect cores. Also, the quadrupolar structure of Fig.~\ref{fig:eta_psi_field_during_nucleation}(a) {\em prior} nucleation reflects the Burgers vectors of the dislocation pair to be nucleated. More quantitatively, Fig.~\ref{fig:eta_psi_field_during_nucleation}(c) shows the evolution of the maximal value of $\eta^\psi$ in time and upon increasing $\sigma_{0}$ quasistatically (shown by the left y-axis). 
The point at which $\sigma_0$ attains the critical value for nucleation is marked by the vertical dashed line at $t=2050$ and the dislocations become distinct at $t\approx 2300$. 
We observe that max($\eta^\psi$) rises before the dislocations become distinct. 
Prior to nucleation, the crystal lattice is elastically loaded with quasistatic increase of $\sigma_0$. Post nucleation, the external stress $\sigma_0$ remains constant (corresponding to a plateau in the $\sigma_0$ curve), while the crystal lattice evolves diffusively in time. 

\begin{figure}
    \centering
    \includegraphics[width=0.35\textwidth]{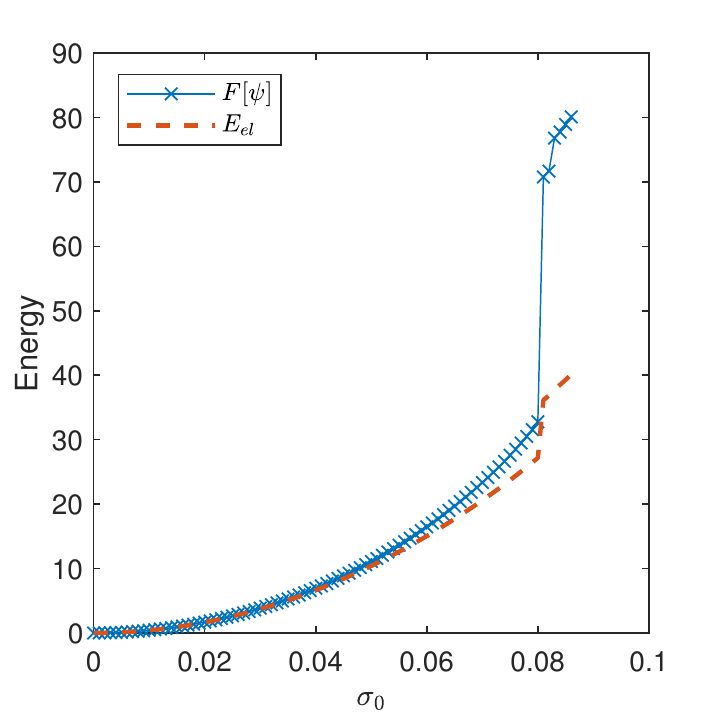} 
    \caption{Total free energy $\mathcal{F}$ and elastic energy $E_{el}$ as function of $\sigma_0$.} 
    \label{fig:free_energy_during_nucleation}
\end{figure}

Fig. \ref{fig:free_energy_during_nucleation} further shows the corresponding change in the PFC free energy $\mathcal{F}$ upon increasing $\sigma_{0}$, together with the elastic energy defined as $ E_{el} = \frac{1}{2} \int d\vec r \sigma_{ij}^\psi e_{ij}^\psi$. Note that despite the purely diffusive dynamics obeyed by $\psi$, the lattice is capable of storing (reversible) elastic energy upon increasing the value of $\sigma_{0}$, as seen previously in Fig. \ref{fig:eta_psi_field_during_nucleation}(c). This reversible evolution is enabled through the compatible distortion added to the field $\psi$ to preserve elastic equilibrium. As the nucleation event is reached, the phase field energy exhibits a large discontinuity at the same value of $\sigma_{0}$ that corresponds to the dashed line in Fig.~\ref{fig:eta_psi_field_during_nucleation}(c).

%
\begin{figure}[t]
    \centering
     \includegraphics[width=0.5\textwidth]{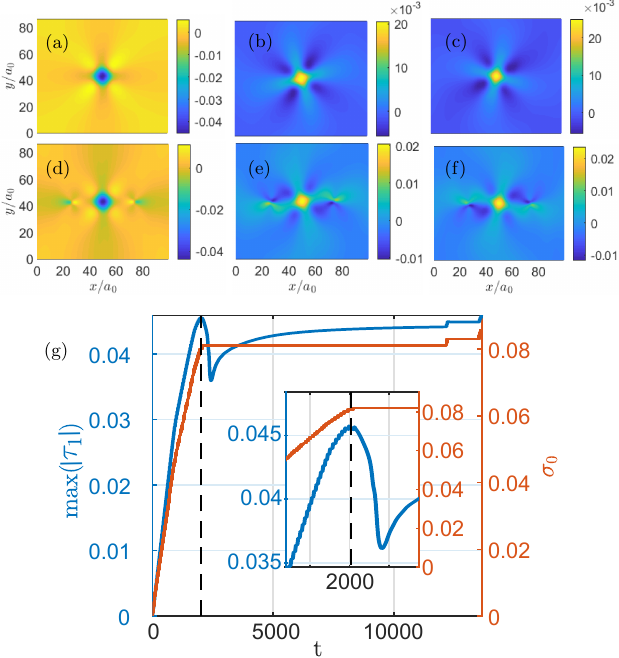} 
\caption{The resolved shear stresses just prior to and after the nucleation event. a)-c) $\tau_1$, $\tau_2$, $\tau_3$ at $t = 2050$ ($\sigma_0=0.080$), respectively, d)-f)  $\tau_1$, $\tau_2$, $\tau_3$ at $t = 12170$ ($\sigma_0=0.081$), respectively. g) $\max(|\tau_1|)$ as function of time $t$ during the nucleation event.}
    \label{fig:tau1tau2tau3nucleation_criteria}
\end{figure}
%
\begin{figure*}
\centering
   \includegraphics[width=0.9\textwidth]{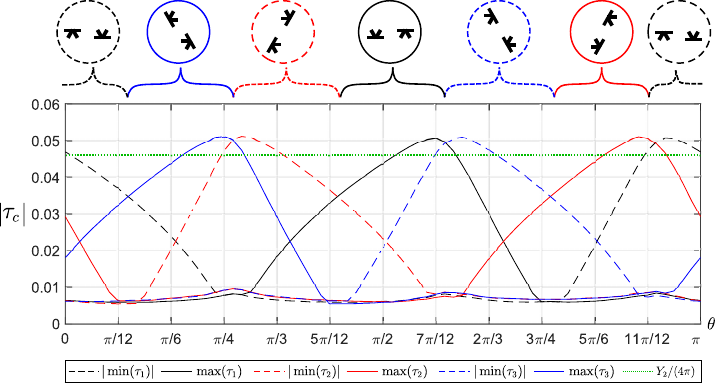}    
    \caption{The value of $|\tau_c|$ at nucleation as a function of the rotation angle $\theta$ of the externally imposed stress $\sigma_{ij}^{\ext}(\theta)$. 
    The top row shows the type of dislocation dipole that nucleates.
    The resolved stress along the slip plane with the largest value determines the type of dislocation dipole to nucleate. }
    \label{fig:max_tau_during_nucleation_as_function_of_external_stress_rotation}
\end{figure*}
%
For this simple 2d setup, it is possible to predict the critical stress for nucleation from the value of the resolved shear stress along each slip plane, in analogy with the classical Schmid criterion. For a given stress $\sigma_{ij}$, the resolved shear stress $\tau_{\vec a,\vec n}$ on a slip plane defined by the normal unit vector $\vec n$ along a direction in the slip plane given by the unit vector $\vec a$ is $ \tau_{\vec a,\vec n} = a_{i} \sigma_{ij} n_j$. In 2D, $\vec n$ is determined up to a sign by $n_i = \e_{ij} a_j$, and for the hexagonal symmetry, there are three slip planes defined by lattice vectors $\vec a_1=[1,0], \vec a_2 = [1/2,\sqrt 3/2]$, and $\vec a_3 = [-1/2,\sqrt 3/2]$. One thus considers three different scalar fields $\tau_1,\tau_2$, and $ \tau_3$, which are the resolved shear stresses along the slip directions corresponding to $\vec a_1, \vec a_2$ and $\vec a_3$, respectively. Figs. \ref{fig:tau1tau2tau3nucleation_criteria}(a-f) show the fields $\tau_1, \tau_2, \tau_3$ right before and after nucleation. The resolved shear stress is largest along the $\vec a_1$ direction, the slip plane along which the dislocation pair nucleates, and is centered at the origin, the nucleation site. The other two resolved stresses remain small during nucleation. The change in the largest resolved stress $\tau_1$ is shown in Fig.~\ref{fig:tau1tau2tau3nucleation_criteria}(g), using the same coordinates as in Fig. \ref{fig:eta_psi_field_during_nucleation}(c). Nucleation initiates (vertical dashed line in the figure) when the resolved shear stress approaches the critical value of $| \tau_c | = 0.046$, followed by a small drop, and then a slow rise as the newly nucleated dislocation dipole moves away from the center region. Notice that this value of $\tau_c$ at the moment of nucleation is smaller than the external shear stress $\sigma_{xy}^{ext} = 0.080$. This is because at mechanical equilibrium, the two stresses are equal only up to a divergence-free term. The critical value of the resolved stress $ \tau_c $ can be estimated as follows: Consider an otherwise perfect lattice with a bound dislocation pair of opposite Burgers vectors. The force acting on the dislocations (in opposite directions) because of the external stress is the Peach-Koehler force projected on the slip plane defined by $\vec a_k$, and is $ F_k^{PK} = b \tau_{k} = \pm a_0 \tau_k$, for dislocations with Burgers vectors $\vec b=\pm a_0 \vec a_k$. As the two dislocations in the dipole separate at nucleation to become distinct, their mutual elastic interaction results in an attractive force. If both dislocations are on the $x$ axis, this force is \cite{CMPprinciples,re:perreault16} $ | f_{x} | = Y_{2}b^{2}/(4 \pi d)$, where $Y_{2} = 4\mu(\lambda + \mu)/(\lambda + 2 \mu)$ is the 2D Young modulus, and $d$ the dislocation separation. We estimate $\tau_{c}$ as the applied stress for which the resulting Peach-Kohler force on one dislocation equals the force from the other dislocation when the separation is one lattice constant. We find that $\tau_{c} = Y_{2}/(4 \pi)$. Using the numerical values of $\mu = \lambda = 0.2271$, $ Y_{2}/(4 \pi) = 0.048$, which is in close agreement with the observed value of $|\tau_c|=0.046$.

In order to further test the nucleation criterion, we have performed additional calculations in which the imposed stress $\sigma_{ij}^{ext}$ is rotated relative to the lattice, $\sigma_{ij}^{ext}(\theta) = R^{(\theta)}_{ki} \sigma_{kl}^{ext} R^{(\theta)}_{lj}$, where $R^{(\theta)}_{ij}$ is the standard rotation matrix in 2D, and $\theta$ is the rotation angle.
Fig. \ref{fig:max_tau_during_nucleation_as_function_of_external_stress_rotation} shows the maximal resolved stress at nucleation along the three lattice directions as a function of $\theta$. Since $\sigma_{ij}^{ext}$ is invariant under a rotation of $\pi$, $\sigma_{ij}^{ext}(x) = \sigma_{ij}^{ext}(\pi+x)$, we show only values ranging from $\theta=0$ to $\pi$. The figure shows that the resolved stress consistently predicts the type of dislocation dipole to nucleate, but the value of the critical resolved stress depends on $\theta$, and is in general lower than $Y_{2}/(4 \pi)$. The discrepancy is likely due to anisotropic contributions to lattice distortions at the length scale of the core which are not described by isotropic linear elasticity.

\section{3D Incompatibility field}

The simplest example of a 2D hexagonal lattice has only point edge dislocations and is described by isotropic elasticity. However, more realistic crystal lattices have more complex loop defects and lattice anisotropy, where Schimd like criterion might not readily apply. Therefore, it is important to understand how the incompatibility field applies to 3D and behaves near a nucleation event. Here we derive the incompatibility field from the $\psi$ density field corresponding to a bcc lattice in 3D and visualize it for a dislocation that is seeded into an otherwise perfect crystal. 
Since the incompatibility field is determined by the topology of the system, it accurately locates the dislocation strings and provides a powerful tool to visualize mixed edge/screw dislocation loops. 

\begin{figure*}
    \centering
    \includegraphics[scale=1]{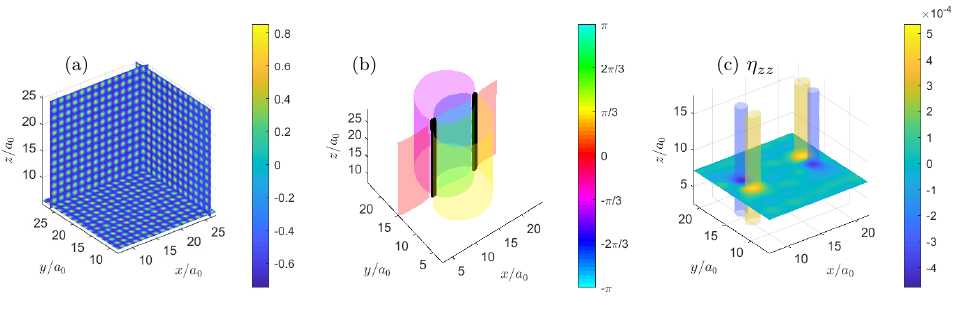}
    \caption{
    a) 2D slices of the field $\psi$ showing edge dislocations at $[x,y] = [10a_0,15a_0]$ and $[x,y] = [20a_0,15 a_0]$.
    b) The amplitude $A_1$ of the two dislocation lines in the PFC model. The black subvolume indicates the zeros of the amplitudes and thus the position of the dislocation lines while the colormap gives the complex argument. c) The $\eta_{zz}^\psi$-component of the incompatibility of dislocation lines in the PFC model. 
    The other components of the incompatibility tensor are small relative to this component (see text).}
    \label{fig:PFC_of_dislocation_lines_in_xyplane}
\end{figure*}

For a suitable range of parameters $\bar \psi$ and $r$, the equilibrium state that minimizes the free energy functional in Eq. \ref{eq:free_energy_of_PFC_model} is given by a body-centered-cubic (bcc) lattice in 3D. The corresponding reciprocal lattice vectors lie on a face-centered-cubic (fcc) lattice with a lattice constant of unity. We choose as parameters values $\bar \psi = -0.371$, $r=-0.4$ for the results presented below. The amplitude of the reciprocal modes in equilibrium is  $A_{0} = -\frac{2}{15}\bar \psi + \frac{1}{15}\sqrt{-5 r - 11\bar \psi} = 0.2139$ \cite{wuPhasefieldCrystalModeling2007}. The orientation of the lattice is chosen by defining the following set of reciprocal lattice vectors of unit length: $\vec q_1 = [1,1,0]/\sqrt 2$, $\vec q_2 = [1, 0 , 1]/\sqrt 2$, $ \vec q_3 = [0  , 1, 1]/\sqrt 2$, $ \vec q_4 = \vec q_1 - \vec q_3$, $\vec q_5 = \vec q_2 - \vec q_3$, $\vec q_6 = \vec q_1 - \vec q_2$. A cubic computational domain is considered with $30 \times 30 \times 30$ bcc unit cells of length $a_0=2\pi$ with grid spacings $\Delta x = \Delta y = \Delta z= a_0/4$.

We first examine a configuration with two dislocation lines added to the phase field by multiplying the initially constant amplitudes $A_{\vec q_n}$ of the PFC by phase factors $e^{s_{n}i\theta}$ corresponding to (i), a pure edge dislocation with Burgers vector $\vec b_1 = \vec e_x$ and constant tangent line $\vec l_1=\vec e_z$ at $[x_1,y_1]=[20a_0,15a_0]$ and, (ii), a pure edge dislocation with Burgers vector $\vec b_2 = -\vec e_x$ and constant tangent line $\vec l_1=\vec e_z$ at $[x_1,y_1]=[10a_0,15a_0]$. Here $\theta$ is the angle in the $xy$-plane relative to the $x$-plane, and $s_n$ is the charge of the dislocation, calculated as in Ref. \cite{skaugen2018dislocation}. The PFC is subsequently prepared, in the one-mode approximation, as $\psi = \psi_0 +  \sum_{n=1}^6 [A_{\vec q_n}(\vec r) e^{i\vec q_n\cdot \vec r} + c.c.]$ and allowed to evolve diffusively for few time steps to regularize the dislocation core.  The stress tensor is calculated according to Eq. \ref{eq:stress_tensor}, and the strain $e_{ij}^\psi$ is found by inverting the stress tensor according to linear elasticity, using the (anisotropic) elastic constants of the bcc lattice given in Ref. \cite{wuPhasefieldCrystalModeling2007}.

The incompatibility is now a rank-2 tensor with components given by $ \eta_{ab}^\psi = \e_{aci} \e_{bdj} \d_{cd} e^\psi_{i j}.$ \cite{re:kronerContinuumTheoryDefects1981}.
Fig. \ref{fig:PFC_of_dislocation_lines_in_xyplane}(a) shows 2D slices of the the PFC after relaxation, with the complex amplitude $A_{\vec q_1}$ determined by amplitude demodulation of the phase field \cite{skaugenSeparationElasticPlastic2018} in Fig. \ref{fig:PFC_of_dislocation_lines_in_xyplane}(b), and the largest component $\eta_{zz}^\psi$ of the incompatibility tensor, Fig. \ref{fig:PFC_of_dislocation_lines_in_xyplane}(c). The figure demonstrates how the core of the dislocation lines become zeros of the complex amplitudes, with a phase discontinuity of $2\pi$ going around a dislocation line. The incompatibility tensor in terms of the dislocation density tensor $\a_{ij}$ is given by $\eta_{ik} = (\e_{ipl} \d_p \a_{kl} + \e_{kpl} \d_p \a_{il})$ \cite{kosevichCrystalDislocationsTheory1979}.
For a straight dislocation line with Burgers vector $\vec b= a_0 \vec e_x$ and tangential vector $\vec l = \vec e_z$ as illustrated in Fig. \ref{fig:PFC_of_dislocation_lines_in_xyplane}(b) in black lines, the dislocation density tensor is given by its only non-zero component $\alpha_{zx}$, which gives
$\eta_{xx} = \eta_{yy} = \eta_{xy} = \eta_{xz} = \eta_{yz} = 0$ and $\eta_{zz} = -\d_y \a_{zx}$, which is shown in Fig. \ref{fig:PFC_of_dislocation_lines_in_xyplane}(c). 
Thus, in this case, $-\eta_{zz}$ is the $y$-component of the gradient of the dislocation density, which explains its dipole structure. 
The spatial extent of $\eta_{zz}$ around the dislocation line gives a measure of the spatial smoothing of the dislocation core \cite{PFCamplitudesMarco20}.
This configuration is the straightforward extension of the 2D edge dislocations of Fig. \ref{fig:PFC_during_nucleation}(b) to 3D. This explains the similarity between the 2D slice of Fig. \ref{fig:PFC_of_dislocation_lines_in_xyplane}(c) to Fig. \ref{fig:eta_psi_field_during_nucleation}(b).

\begin{figure*}
    \centering
    \includegraphics[scale=1]{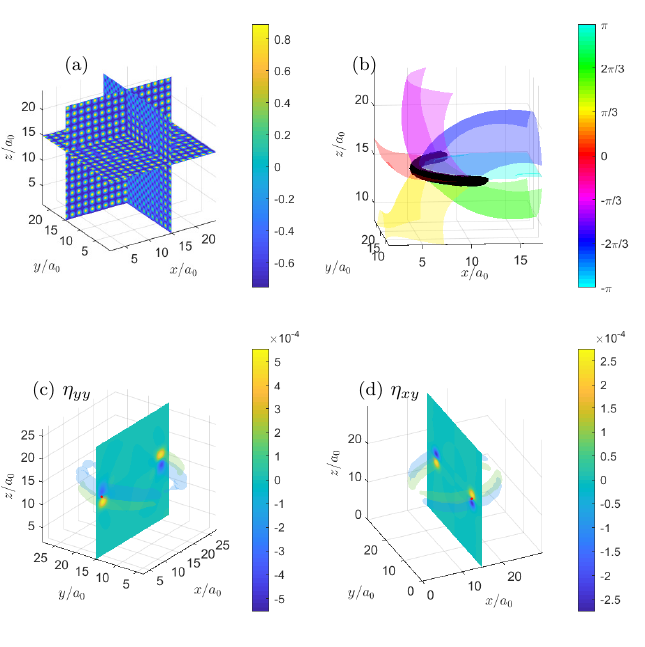}
    \caption{
    a) 2D slices of the field $\psi$ showing the mixed types of dislocations that appear for a dislocation loop.
    b) The amplitude $A_1$ of the dislocation loop in the PFC model. The black subvolume indicates the zeros of the amplitudes and thus the position of the dislocation lines while the colormap gives the argument. 
    c)-d) Two components of the incompatibility tensor $\eta_{ij}^\psi$.
    The red dots indicate positions at which the incompatibility field assumes a dipole structure similar to that of Fig. \ref{fig:PFC_of_dislocation_lines_in_xyplane}(c) as explained in the text.
    }
    \label{fig:PFC_of_dislocation_loop_in_xyplane}
\end{figure*}

In order demonstrate the intrinsic capability of the phase field and its associated incompatibility field to identify dislocations of mixed edge/screw character, we prepare an initial configuration with a dislocation loop. The Burgers vector of the dislocation line is constant and equal to $a_0 \vec e_x$, while the tangent vector $\vec l$ rotates in the $xy$-plane. Since $\vec l$ switches between being parallel to $\vec b$ and perpendicular, this leads to a mixed edge/screw dislocation. 
Fig. \ref{fig:PFC_of_dislocation_loop_in_xyplane}(a) shows 
2D slices of the PFC including the defect after relaxation, and the amplitude $A_{\vec q_1}$ of the first reciprocal lattice vector is shown in Fig. \ref{fig:PFC_of_dislocation_loop_in_xyplane}(b).
For an ideal dislocation loop with Burgers vector $\vec b = a_0 \vec e_x$ along the dislocation loop purely in the $xy$-plane, we only obtain a contribution to the dislocation density tensor from $\a_{ix}$ with $i\neq z$.
To see how the incompatibility field calculated from $\psi$ captures the nature of the defects, consider the portion of the dislocation loop at $\vec r = [5,15,15]$ (red dot in Fig. \ref{fig:PFC_of_dislocation_loop_in_xyplane}(c)).
Here we have $\vec l = -\vec e_y$ which gives $\alpha_{yx}$ as the non-zero component of the dislocation density tensor. 
We get $\eta_{yy} = \d_z \a_{yx}$, the $z$-component of the gradient of the dislocation density, and an identical dipole structure as Fig. \ref{fig:PFC_of_dislocation_lines_in_xyplane}(c) appears, this time in the $z$-direction. 
Similarly at $\vec r = [15,5,15]$ (red dot in Fig. \ref{fig:PFC_of_dislocation_loop_in_xyplane}(d)), the non-zero component of the dislocation density tensor is $\a_{xx}$ from which $\eta_{xy} =  \d_z \a_{xx}/2$ we recover the dipole structure of Fig \ref{fig:PFC_of_dislocation_lines_in_xyplane}(c) and \ref{fig:PFC_of_dislocation_loop_in_xyplane}(c), its magnitude halved due to the factor of $1/2$ in $\eta_{xy}$.

We have exemplified here how the phase field $\psi$ and its associated incompatibility tensor can correctly describe any dislocation string or loop in a given bcc lattice. In further studies, this formalism can be further generalized to other lattice symmetries and also coupled with the evolution of the $\psi$-field to study the dynamics and nucleation of dislocation strings. 

\section{Conclusion}

We have shown for the case of a 2D hexagonal lattice that the incompatibility field $\eta^\psi$ derived from the phase field is a very sensitive diagnostic of the nucleation of a dislocation dipole, and that it signals the nucleation event prior to the formation of a stable topological dipole. The symmetry of the field $\eta^\psi$ prior to nucleation also gives the direction of the Burgers vectors of the defect pair about to nucleate. By examining the distribution of the resolved stress for a hexagonal lattice, we have also found it to be a good indicator of nucleation. Furthermore, a balance between the Peach-Kohler force on either one of the defects of the dipole and their mutual elastic interaction force allows a prediction of the resolved critical stress at nucleation that agrees well with the numerical results. 

While our results serve to extend those of macroscopic plasticity by allowing the direct observation of the incompatibility field and its evolution under an applied stress, the conclusion that a Schimd like criterion identifies the nucleation event is in contrast with several existing Molecular Dynamics simulations. Some of these simulations show that the resolved stress does not predict the location nor type of dislocations to nucleate. Instead, it is generally observed that the nominal extent of the nucleation region is very large, with a complex network of stacking faults, partial dislocations, and other significant sources of lattice distortion. These results would imply that the nucleation path in 3D configuration space can be much more complex than in our 2D configuration, with possibly multiple competing trajectories that depend on details such as boundaries or applied stress protocols. 

Our results indicate that the Phase Field Crystal model provides adequate control over configurations and applied stresses around the nucleation threshold, and hence is a suitable platform for testing nucleation criteria. The model offers the necessary separation between length scales, eliminates fluctuations of thermal origin, and allows the computation of internally generated stress that contribute to lattice incompatibility, and ultimately to nucleation. This bypasses the need for extensive averaging of Molecular Dynamics trajectories along paths in configuration space in which fluctuations are very small.

Finally, we present results for a 3D bcc lattice to show that the phase field can be used to describe dislocations in this lattice, and that it offers the possibility of computing the incompatibility tensor directly from the phase field for arbitrary, non-equilibrium configurations.

\begin{acknowledgments}
We thank Amit Acharya, Kristian Olsen and Jonas R\o nning for many stimulating discussions. The research of J.V. has been supported by the National Science Foundation, Grant No. DMR-1838977. L.A. acknowledges support from the Kavli Institute for Theoretical Physics through the National Science Foundation under Grant No. NSF PHY-1748958. 
\end{acknowledgments}

\bibliographystyle{apsrev4-1}
\bibliography{bibliography}

\end{document}